\documentclass[twocolumn,secnumarabic,amssymb, nobibnotes, aps, pra]{revtex4-1}

\usepackage{framed}
\usepackage{amsfonts}
\usepackage{amsmath}
\usepackage{graphicx}
\usepackage{color}

\usepackage[hidelinks]{hyperref}
\usepackage{xcolor}
\hypersetup{
	colorlinks,
	linkcolor={blue!80},
	citecolor={blue},
	urlcolor={blue!80!black}
}

\begin{document}
\title{Dissipative localized states and breathers in phase mismatched singly resonant optical parametric oscillators: Bifurcation structure and stability}
\author{P. Parra-Rivas, C. Mas Arab\'i, and F. Leo}

\affiliation{OPERA-photonics, Universit\'e libre de Bruxelles, 50 Avenue F. D. Roosevelt, CP 194/5, B-1050 Bruxelles, Belgium\\
}

\date{\today}


\begin{abstract}
We study the emergence of dissipative localized states in phase mismatched singly resonant optical parametric oscillators.
 These states arise in two different bistable configurations due to the locking of fronts waves connecting the two coexisting states.  In one of these configurations the bistabiity is mediated by the coexistence of two uniform states. Here the localized states are organized in a collapsed snaking bifurcation structure. Moreover, these states undergo oscillatory instabilities which lead to a breathing behavior. When the the bistability is related with the coexistence of an uniform state and a spatially periodic pattern, localized states  are organized in a bifurcation structure similar to the standard homoclinic snaking. Performing an exhaustive bifurcation analysis, we characterize in detail the previous structures, their linear stability and the modification of their dynamics as a function of the control parameters of the system.    
\end{abstract}
\maketitle

\section{Introduction}
Spatial localization is a fascinating emergent phenomenon appearing in a large variety of extended nonlinear natural systems out of the thermodynamic equilibrium, also known as dissipative systems \cite{descalzi_localized_2011,knobloch_spatial_2015}. Localized coherent structures (LSs) may appear spontaneously due to a continuous balance between spatial coupling and nonlinearity on one hand, and a continuous exchange of either energy or matter with the surrounding medium on the other hand \cite{akhmediev_dissipative_2005}. Dissipative LSs have been found in the context of solid mechanics, population dynamics, biology and optics, to cite only a few \cite{descalzi_localized_2011,knobloch_spatial_2015}. 
The formation of LSs is generally related with the coexistence of different stable states, domains or phases within the same interval of the control parameters of the system, in what is called a {\it bistable} regime. In this context, front waves or domain walls connecting such states can arise \cite{pomeau_front_1986}, interact, and lock \cite{pomeau_front_1986,coullet_nature_1987}, leading to a plethora of LSs of different morphology and extension  \cite{pomeau_front_1986,coullet_nature_1987,woods_heteroclinic_1999,coullet_stable_2000,coullet_localized_2002,knobloch_homoclinic_2005,beck_snakes_2009}. 

One interesting example supporting this type of states are driven optical cavities with Kerr (cubic) nonlinearity, where light can be trapped and propagates indefinitely. The first experimental confirmation of cavity LSs is due to Barland {\it et al.} who in 2002 generated two-dimensional LSs in the transverse direction of a semiconductor diffractive microcavity \cite{barland_cavity_2002}, the so called {\it spatial cavity solitons}. 
Later, in 2010, one-dimensional LSs were also shown experimentally in fiber cavities by Leo {\it et al.} \cite{leo_temporal_2010}. In this case, LSs form along the propagation direction and are referred to as {\it temporal cavity solitons}. Since then, LSs have been found in other type of Kerr dispersive cavities and in different operational regimes \cite{herr_temporal_2014,xue_mode-locked_2015}. In dispersive cavities, LSs
have been proposed for different technological applications including all-optical buffering \cite{leo_temporal_2010} and frequency comb generation \cite{herr_temporal_2014,brasch_photonic_2016}.

Another example of optical systems potentially supporting LSs are pure quadratic nonlinear cavities.  During the '90s and early '00s a lot effort was dedicated to the theoretical characterization of LSs in diffractive quadratic cavities \cite{staliunas_localized_1997,longhi_localized_1997,trillo_stable_1997,staliunas_spatial-localized_1998,oppo_domain_1999,oppo_characterization_2001,rabbiosi_new_2003,etrich_solitary_1997,Trillo:98}. 
Recently, pure quadratic dispersive cavities attracted renewed attention due to their multiple advantages for frequency combs generation \cite{ulvila_frequency_2013,leo_walk-off-induced_2016,leo_frequency-comb_2016,hansson_singly_2017,mosca_frequency_2017,mosca_modulation_2018}. Since then, different theoretical studies have tackled the investigation of light localization in either cavity enhanced second-harmonic generation \cite{hansson_quadratic_2018,villois_frequency_2019,arabi_localized_2020} or in degenerate optical parametric oscillators (DOPOs) \cite{parra-rivas_frequency_2019,parra-rivas_localized_2019,parra-rivas_parametric_2020,nie_quadratic_2020,nie_quadratic_2020-2}. 

Temporal LSs have recently attracted attention in the context of competing nonlinearities.  Specifically,  Kerr solitons have been shown to form in quadratic optical parametric oscillations \cite{bruch_pockels_2021,NieXieHuang2021,englebert_parametrically_2021}. These (${\rm sech}$ shape) LSs correspond to {\it parametrically driven} Kerr solitons and have been proposed as key elements for random number generation and Ising-machines. 


Here, we investigate the formation of parametrically driven LSs in phase mismatched singly resonant OPOs.
The effective Kerr nonlinearity, inherent to mismatched quadratic processes \cite{DeSalvo:92}, supports the formation of LSs which are absent in the phase-matched configuration.
The presence of {\rm sech} pulses in phase-mismatched OPOs have already been investigated,  both in the singly resonant \cite{nie_quadratic_2020,nie_quadratic_2020-2} and in the doubly resonant \cite{longhi_localized_1997,Trillo:98} configurations.
In this paper, we focus on the former and investigate the range of existence and stability of these solitons as well as uncover the presence of other LSs.

The paper is organized as follows. In Sec.~\ref{sec:1} we introduce the mean-field model describing dispersive singly resonant DOPOs. Later, in Sec.~\ref{sec:2} the uniform or homogeneous steady state of the system is presented and its spatio-temporal linear stability is analyzed. Doing so, we can determine the existence of two different scenarios, namely the uniform-bistable and Turing-bistable regimes. Section~\ref{sec:3} is devoted to the computation of a weakly nonlinear localized solution close to a relevant bifurcation of the trivial homogeneous state. This solution will be then used as initial guess in a numerical continuation procedure to compute the bifurcation structure of LSs in the highly nonlinear regime (Secs.~\ref{sec:4} and \ref{sec:5}). In Sec.~\ref{sec:4} such structure is computed in a uniform-bistable configuration, and Sec.~\ref{sec:4b} is devoted to the study of the oscillatory dynamics undergone by the LSs in this regime. Later, in Sec.~\ref{sec:5} we focus on the Turing-bistable scenario. After that, in Sec.~\ref{sec:6} we study how spatial symmetry breaking can impact the dynamics and bifurcation structure of such states. Finally, in Sec.~\ref{sec:8} we conclude with the general discussion and the conclusions.

\section{The mean-field model}\label{sec:1}

\begin{figure}[t]
	\centering
	\includegraphics[scale=1]{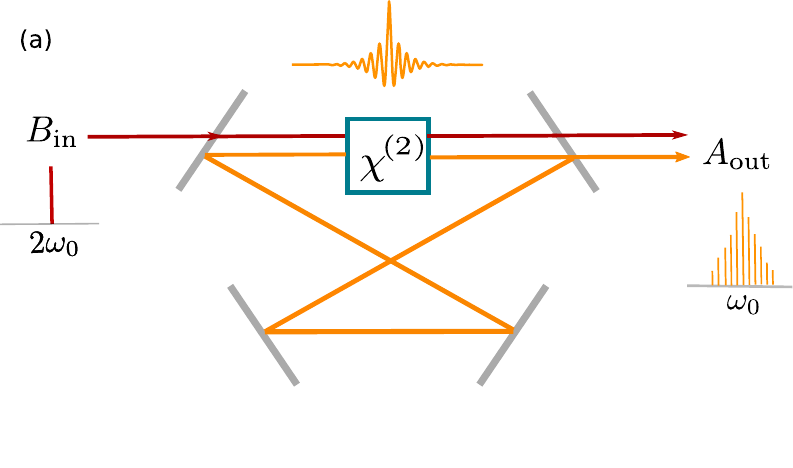}
	\includegraphics[scale=1]{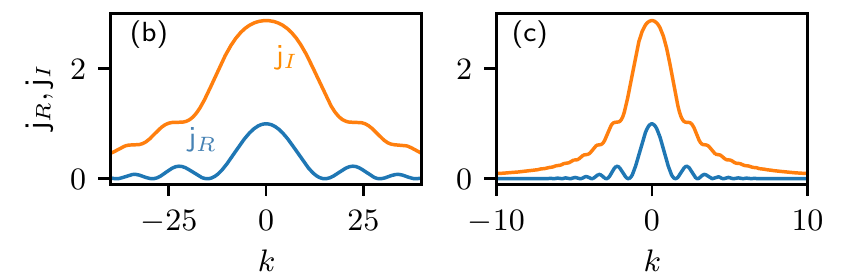}
	\caption{Panel (a) shows an schematic example of a singly resonant DOPO system. The cavity contains a nonlinear quadratic $(\chi^{(2)})$ medium and is driven by a continuous-wave field $B_{in}$ at frequency $2\omega_0$. Here only the generated field $A$ at frequency $\omega_0$ resonates, while the pump field $B$ leaves the cavity at each round-trip.
		Panels (b) and (c) show $\mathsf{j}_R(k)$ and $\mathsf{j}_I(k)$ in the absence of walk-off ($d=0$) and $\varrho=-4$, for $\eta_2=0.01$, and $\eta_2=1$, respectively.}
	\label{fig0}
\end{figure}

In singly resonant DOPOs a continuous-wave field $B_{in}$ at frequency $2\omega_0$ (the pump field) is injected in the cavity (see Fig.~\ref{fig0}). Within the cavity, this field interact nonlinearly with the quadratic $(\chi^{(2)})$ material and triggers the generation of a {\it signal} field at frequency $\omega_0$, due to a parametric frequency down conversion process. In contrast to doubly resonant cavities where both fields resonate, here the pump is extracted at every round-trip, and only the signal field resonates.

In the mean-field approximation, the Ikeda map describing the cavity can be reduced to the following dimensionless partial differential equation with nonlocal nonlinearity \cite{mosca_modulation_2018}
\begin{equation}\label{normalized}
\partial_t {\mathsf A}=-(1+i\Delta_1)\mathsf{A}-i\eta_1\partial_{x}^2\mathsf{A}-\bar{\mathsf{A}}(\mathsf{A}^2\otimes\mathsf{J}) +S\bar{\mathsf{A}},
\end{equation}
where $\mathsf{A}$ is the slowing varying envelope of the signal electric field circulating in the cavity and $\bar{A}$ its complex conjugate, $\Delta_1$ is the normalized phase detuning from the closest cavity resonance, $\eta_1$ represents the normalized group velocity dispersion of $\mathsf{A}$ at $\omega_0$ and $S$ is the normalized amplitude of the driving field. The term $\mathsf{A}^2\otimes\mathsf{J}$ represents the nonlocal nonlinearity defined through the convolution ($\otimes$) between $\mathsf{A}^2$ and the nonlocal kernel
\begin{equation}
\mathsf{J}(x)=\frac{1}{2\pi}\int_{-\infty}^{\infty}\mathsf{j}(k)e^{-ik x}dx.
\end{equation}
The Fourier transform of $\mathsf{J}$, namely $\mathcal{F}[\mathsf{J}](k)\equiv\mathsf{j}(k)=\mathsf{j}_R(k)+i\mathsf{j}_I(k)$ is defined through the expressions
\begin{figure*}[t]
	\centering
	\includegraphics[scale=1]{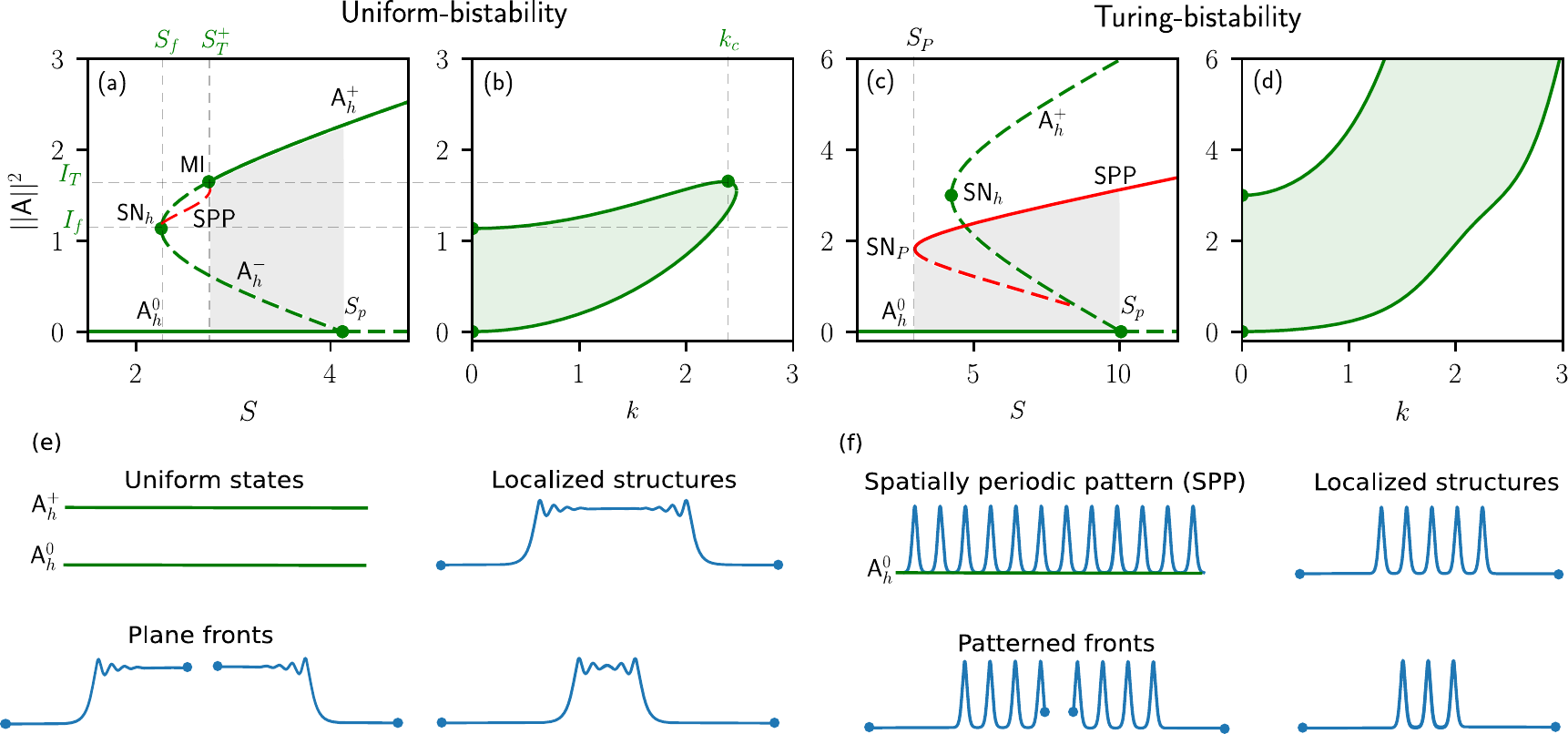}	
	\caption{Panels (a) and (b) show respectively the homogeneous states $\mathsf{A}_h$ and their stability in a subcritical regime, and its associated MIC for $(\Delta_1,\varrho,\eta_2)=(-4,-4,0.01)$. This configuration shows uniform-bistability in the interval $S_T<S<S_p$ (see shadowed area). The maximum of the MIC corresponds to the Turing instability or MI occurring at $(I_h,S)=(I_T,S_T)$. Stable (unstable) branches are plotted with solid (dashed) lines.
		Panels (c) and (d) a subcritical configuration for $(\Delta_1,\varrho,\eta_2)=(-10,-4,1)$. Here, in contrast, ${\mathsf A}_h^+$ is always modulationally unstable [see panel (c)] as shown in its corresponding MIC [see panel (d)]. In this configuration uniform-bistability is not present. However, different Turing pattern may bifurcate from branching points all along $\mathsf{A}^\pm_h$, as the one shown in red in (c). The interval $S_{P}<S<S_p$ defines the Turing-bistability region (see shadowed area).     }
	\label{fig1}
\end{figure*}
\begin{subequations}
	\begin{equation}
	\mathsf{j}_R(k)=\frac{{\rm sinc}^2(Z(k)/2)}{2a(\varrho)},
	\end{equation}
	\begin{equation}
	\mathsf{j}_I(k)=\frac{{\rm sinc}(Z(k))-1}{a(\varrho)Z(k)},
	\end{equation}
\end{subequations}
where
\begin{align}
Z(k)\equiv\varrho-dk-\eta_2 k^2,&&a(\varrho)\equiv\frac{1}{2}{\rm sinc}^2(\varrho/2),
\end{align}
and $\varrho$, $d$ and $\eta_2$ are the normalized phase mismatch, group velocity mismatch or walk-off, and group velocity dispersion of the pump field. Here, $\mathsf{j}_R(k)$ mirror the dispersive two-photon absorption, while  the contribution related with $\mathsf{j}_I(k)$ produces a phase-shift, similar to the Kerr effect \cite{nie_quadratic_2020}. More details about the model and the normalization considered here are presented in Appendix~\ref{sec:A}.

In what follows, we first study the features and dynamics of LSs in the absence of walk-off ($d=0$), and later, in Sec.~\ref{sec:6}, we elucidate the implications that a weak walk-off may have on our results. 

For $d=0$, the nonlocal response $\mathsf{J}$ depends only on $\varrho$ and $\eta_2$, and two different regimes can be identified depending on the sign of $\varrho\eta_2$ as reported in \cite{nie_quadratic_2020}. In this work, we focus on the regime $\varrho\eta_2<0$, where $\mathsf{j}_R(k)$ and $\mathsf{j}_I(k)$ vary periodically with $\varrho$, as shown in \cite{nie_quadratic_2020}. Therefore, we fix 
$\varrho=-4\approx-1.27\pi$
 and $\eta_{2}>0$. Figure~\ref{fig0}(b),(c) illustrates the shape of $\mathsf{j}_R(k)$ and $\mathsf{j}_I(k)$ for different values of $\eta_2$.


In most of this work we deal with time-independent states, in particular LSs, satisfying 
\begin{equation}\label{sta_normalized}
 -(1+i\Delta_1)\mathsf{A}-i\eta_1\partial_{x}^2\mathsf{A}-\bar{\mathsf{A}}(\mathsf{A}^2\otimes\mathsf{J}) +S\bar{\mathsf{A}}=0,
\end{equation}
which is formally equivalent to the model describing time-independent states in doubly resonant DOPO \cite{parra-rivas_frequency_2019,parra-rivas_localized_2019}. For $\mathsf{j}_R(k)$ and $\mathsf{j}_I(k)$ with a very low group velocity dispersion at $2\omega_0$, 
 $\mathsf{j}_{R,I}(k)\sim\mathsf{j}_{R,I}(0)$, and Eq.~(\ref{normalized}) can be reduced to a local parametrically forced Ginzburg-Landau equation as shown in Ref.~\cite{parra-rivas_frequency_2019}.

To unveil the dynamics and the bifurcation structure of these states we combine pseudo-spectral exponential time differentiating, or split-step, schemes \cite{montagne_wound-up_1997} for the numerical integration of Eq.~(\ref{normalized}), and numerical
path-continuation algorithms to track the time-independent solutions of Eq.~(\ref{sta_normalized}) \cite{doedel_numerical_1991-1,doedel_numerical_1991}.

The temporal stability of the different time-independent states is obtained by  
solving numerically the eigenvalue problem:
\begin{equation}\label{eigen}
	L\psi=\sigma\psi,
\end{equation}
where $\sigma$ is the eigenvalue associated with the eigenmode $\psi$, and $L$ represents the linear operator obtained from the
linearization of Eq.~(\ref{normalized}) around a given stationary state. Thus, a time-independent state is stable whenever all the eigenvalues satisfy
${\rm Re}[\sigma] < 0$ and unstable otherwise.

The analytical calculations are performed in an infinite domain, while a finite domain of length $l=60$ with periodic boundary conditions is considered for the numerical computations.

\section{Uniform wave state and its temporal stability}\label{sec:2}
The uniform or homogeneous steady state of the system $\mathsf{A}_h$ satisfies $\partial_t\mathsf{A}_h=\partial_x^2\mathsf{A}_h=0$, and therefore is solution of the equation
\begin{equation}
-(1+i\Delta_1)\mathsf{A}_h-(1+i\beta)|\mathsf{A}_h|^2\mathsf{A}_h+S\bar{\mathsf{A}}_h=0,
\end{equation}
with $\beta\equiv\mathsf{j}_I(0)$.
Taking $\mathsf{A}_h=\sqrt{I_h}e^{i\phi}$ with $I_h\equiv |\mathsf{A}_h|^2$,  we get   
$$\left[-(1+i\Delta_1)-(1+i\beta)|\mathsf{A}_h|^2+Se^{-2i\phi}\right]|\mathsf{A}_h|=0,$$
which supports a trivial state solution  $\mathsf{A}_h^0=0$, and non-trivial states $\mathsf{A}_h^\emptyset$ with phase 
\begin{equation}
\phi=\frac{1}{2}{\rm acos}\left(\frac{|\mathsf{A}^\emptyset_h|^2+1}{S}\right).
\end{equation}
The trivial state undergoes a pitchfork bifurcation at 
\begin{equation}
	S_p\equiv \sqrt{1+\Delta_1^2},
\end{equation}
from where the non-trivial state $\mathsf{A}^\emptyset_h$ arises. 

For $\Delta_1<0$, and $|\Delta_1|<1/\beta$, a single non-trivial state 
\begin{equation}
\mathsf{A}^\emptyset_h=r\equiv\frac{\sqrt{(1+\beta^2)S^2-(\Delta_1-\beta)^2}}{1+\beta^2}	
\end{equation}
emerges supercritically from $S_p$. In the following we refer to the point at $\Delta_1=1/\beta$ as the {\it nascent bistability} point. For  $|\Delta_1|>1/\beta$, however, two non-trivial states appear. They read 
\begin{align}
|\mathsf{A}_h^\pm|^2\equiv I_f\pm r, && I_f\equiv-\frac{1+\beta\Delta_1}{1+\beta^2}.	
\end{align}
Here, $\mathsf{A}^-_h$ bifurcates subcritically from $S_p$, and merges with $\mathsf{A}^+_h$ at a fold occurring at $(S,I_h)=(S_f,I_f)$, where
\begin{equation}
S_f\equiv \frac{\Delta_1-\beta}{\sqrt{1+\beta^2}}.
\end{equation}
Here we focus on the subcritical regime ($\Delta_1<0$). Two examples of this configuration are shown in Figs.~\ref{fig1}(a) and \ref{fig1}(c) for $\Delta_1=-4$ and $\Delta_1=-10$ respectively, where 
the energy of $\mathsf{A}$:
$$||\mathsf{A}||^2=\frac{1}{l}\int_{-l/2}^{l/2}|\mathsf{A}(x)|^2dx,$$
is plotted as a function of the amplitude of the driving field $S$. Note that for the homogeneous state  $||\mathsf{A}_h||^2=|\mathsf{A}_h|^2\equiv I_h$.

The next step in our analysis is to determine the linear stability of the previous uniform states against spatiotemporal plane-wave perturbations of the form $e^{\sigma t}\psi_k(x)+c.c.$, where $\sigma$ is the growth rate of the perturbation and $\psi_k$ is the mode $\psi_k(x)\sim e^{ikx}$ associated with the wavenumber $k$. The linear stability analysis around $\mathsf{A}^0_h$ shows that this state undergoes a Turing, or modulational instability (MI), at $S=S_T^0\equiv 0$ whenever $\eta_1\Delta_1>0$ \cite{longhi_localized_1997,parra-rivas_localized_2019}. When the contrary holds,  $\mathsf{A}^0_h$ is always stable up to the pitchfork bifurcation at $S=S_p$, as shown in Figs.~\ref{fig1}(a) and \ref{fig1}(c). We fix $\eta_1=1$ to further investigate this configuration. 

The linear stability of the non-trivial state $\mathsf{A}^\emptyset_h=\mathsf{A}^\pm_h$ is characterized by means of the {\it marginal instability curve} (MIC) obtained from the condition ${\rm Re}[\sigma]=0$ \cite{parra-rivas_localized_2019}. Thus, the MIC is solution of the following quadratic equation 
\begin{equation}
c_2I_h^2+c_1I_h+c_0=0,
\end{equation}
where
\begin{subequations}
	\begin{equation}
	c_2=4(\mathsf{j}_R^2+\mathsf{j}_I^2),
	\end{equation}
	\begin{equation}
	c_1=4(\mathsf{j}_R-(\eta_1k^2-\Delta_1)\mathsf{j}_I),
	\end{equation}
	\begin{equation}
	c_0=\eta_1^2k^4-2\eta_1\Delta_1k^2.
	\end{equation}
\end{subequations}
The MIC defines the band of unstable modes $\psi_k$, and is plotted for two different set of parameters in Figs.~\ref{fig1}(b) and \ref{fig1}(d). 
$\mathsf{A}^\pm_h$ is unstable against a given mode $\psi_k$ if its intensity
$I_h$ lays within the MIC [see shadowed green region in
Figs.~\ref{fig1}(b) and \ref{fig1}(d)], and stable otherwise. In correspondence, Figs.~\ref{fig1}(a) and \ref{fig1}(c) show stable (unstable) uniform states using solid (dashed) lines. The points on the MIC at $k=0$ mark the two stationary instabilities corresponding to the pitchfork bifurcation $S_p$ and the saddle-node bifurcation (SN$_h$), i.e., to the fold occurring at $S_f$.   

Let us first focus on the configuration shown in Figs.~\ref{fig1}(a),(b) for $(\Delta_1,\eta_2)=(-4,0.01)$. Here, $A^-_h$ is always unstable, whereas $A^+_h$ is only unstable between the fold SN$_h$ and the Turing instability occurring at $S_T^{+}$ [see Fig.\ref{fig1}(a)]. At this point, a spatio-temporal perturbation of $\mathsf{A}^+_h$ may slowly evolve to a spatially periodic Turing pattern (SPP) characterized by a critical wavenumber $k_T$. This instability corresponds to the maximum of the MIC occurring at $(k,I_h)=(k_T,I_T)$ as shown in Fig.\ref{fig1}(b). In this case, the Turing pattern arises subcritically from $S_T^+$, but it remains unstable until merging again with $\mathsf{A}^+_h$ in very close to SN$_h$.
The interval $S_T^+<S<S_p$ defines a region of bistability between $\mathsf{A}_h^0$ and $\mathsf{A}_h^+$ [see shadowed area in Fig.~\ref{fig1}(a)]. In what follows we refer to this configuration as {\it uniform-bistability}. In this bistable scenario, {\it plane fronts} or {\it domain walls} can form 
connecting $\mathsf{A}_h^0$ and $\mathsf{A}_h^+$ and vice-versa [see Fig.~\ref{fig1}(e)]. These fronts normally propagate at a constant speed which depends on the control parameter of the system. However, close to the uniform Maxwell point, where the speed cancels out, these fronts can lock one another, leading to the formation LSs of different widths, provided there are damped spatial oscillations ({\it oscillatory tails}) in the front's profile \cite{coullet_localized_2002}. Two examples of these types of LSs are shown in Fig.~\ref{fig1}(e). These LSs undergo a particular bifurcation diagram which is presented in detail in Sec.~\ref{sec:4}. 

Another scenario leading to bistability is the one depicted in Figs.~\ref{fig1}(c),(d) for $(\Delta_1,\eta_2)=(-10,1)$. Here,  $\mathsf{A}_h^-$ and $\mathsf{A}_h^+$ [see Fig.~\ref{fig1}(c)] are always spatiotemporally unstable as shown by the MIC in Fig.~\ref{fig1}(d). For this set of parameters, Turing patterns emerge along $\mathsf{A}_h^-$ and $\mathsf{A}_h^+$. When this patterned state emerges subcritically [see red solid-and-dahsed curve in Fig.~\ref{fig1}(c)], a new bistable configuration appears between $\mathsf{A}_h^0$ and the stable SPP, spanning the parameter region in-between SN$_P$ and $S_p$ [see shadowed area in Fig.~\ref{fig1}(c)]. We refer to this scenario as {\it Turing-} or {\it pattern-bistability}. In this context, {\it patterned fronts} can emerge connecting such stable states [see Fig.~\ref{fig1}(f)], and eventually they can also lock leading to the formation of localized patterns consisting in a slug of SPP embedded in $\mathsf{A}_h^0$ \cite{woods_heteroclinic_1999,coullet_stable_2000}. In Sec.~\ref{sec:5} we will analyze the stability and bifurcation organization of such states.

\section{Weakly nonlinear solutions}\label{sec:3}


In the weakly nonlinear regime, one can analytically compute approximate asymptotic small amplitude LS solutions around some particular bifurcations of the uniform state. In this way, the birth of such states can be understood from a bifurcation perspective. 
\begin{figure}[!h]
	\centering
	\includegraphics[scale=1]{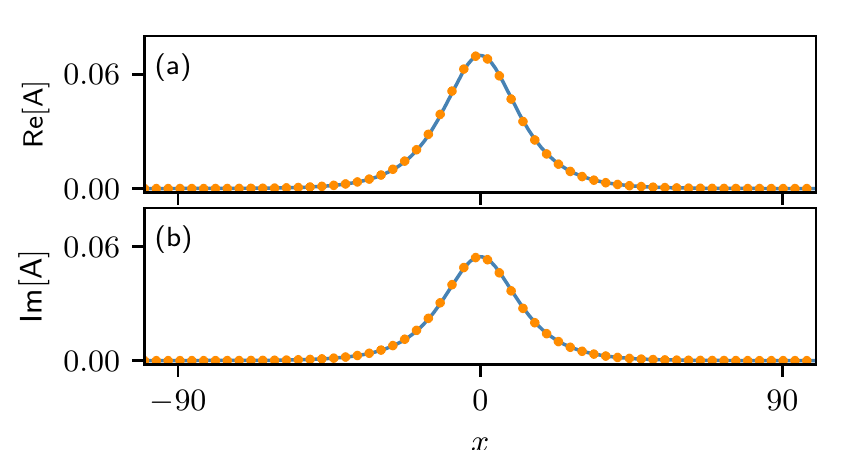}
	\caption{Agreement between the asymptotic weakly nonlinear state (\ref{bump}) computed analytically (see blue solid lines), and the exact numerical solution obtained through a Newton-Raphson solver (orange dots) for $(\Delta_1,\eta_2)=(-4,0.01)$ and $l=200$. Panel (a) and (b) show the agreement of the real and imaginary parts of $\mathsf{A}$ respectively.}
	\label{fig2}
\end{figure}
\begin{figure*}[!t]
	\centering
	\includegraphics[scale=1]{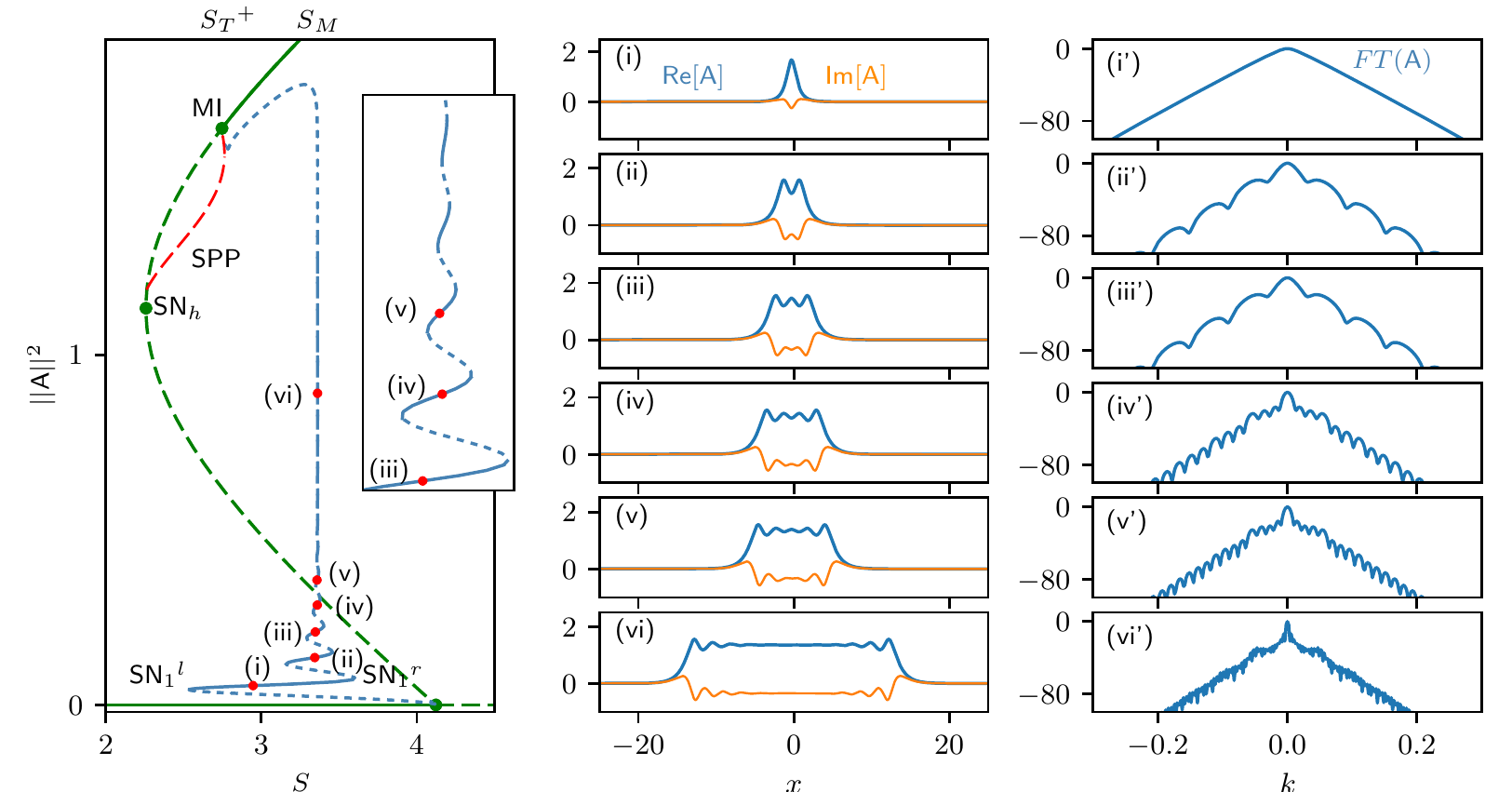}
	\caption{Collapsed homoclinic snaking in the uniform-bistable configuration. This bifurcation curve arises from $S_p$ at $\mathsf{A}_h^0$ snakes in a damped oscillatory manner around the uniform Maxwell point of the system $S_M$. Eventually the diagram connects with subcritical periodic Turing pattern emerging from the MI at $S_T^+$. Panels (i)-(vi) show different LSs solution along this curve. Panels (i')-(vi') represent the frequency spectrum of the LSs. For the computations we have considered $l=60$, and the parameters are fixed to $(\Delta_1,\eta_2)=(-4,0.01)$. }
	\label{fig3}
\end{figure*}
Here, we use multi-scale perturbation theory (see Appendix~\ref{sec:B}) to compute such solutions near the pitchfork bifurcation occurring at $S=S_p$ in an infinite domain system. In the neighborhood of this point, weakly nonlinear states are 
captured by the ansatz:
\begin{equation}
\mathsf{A}(x)-\mathsf{A}_h\sim \epsilon B(X)e^{ik_cx}+c.c.,
\end{equation}
where $\epsilon\ll1$ measures the onset from the bifurcation, $k_c$ is the characteristic wave-number of the marginal mode at the bifurcation, and $B$ is the amplitude or envelope describing a modulation occurring at a larger scale $X=\epsilon^mx$, with the election of $m$ depending on the problem. Close to the pitchfork bifurcation $\mathsf{A}_h=\mathsf{A}_h^0\equiv0$, $k_c=0$ and $m=1$, and the amplitude $B(X)$ is described by the time-independent normal form equation 
\begin{equation}\label{normal_form}
C_1\partial^2_{X}B=\delta B+C_3B^3,
\end{equation}
with the coefficients
\begin{align}
C_1\equiv\frac{-2\eta_1\xi}{1+\xi^2}, && C_3\equiv 2\xi \beta(\varrho)-(\xi^2-1),
\end{align}
where $\xi=\Delta_1/(1-S_p)$, and the bifurcation parameter $\delta$ is defined as $S=S_p+\delta\epsilon^2$.		

In the subcritical regime $(C_3>0)$, this equation supports pulse solutions of the form
$$B(X)= \sqrt{\frac{-2\delta}{C_3}}{{\rm sech}\left(\sqrt{\frac{\delta}{C_1}}X\right)}$$
which leads to the weakly non-linear state 
\begin{equation}\label{bump}
\mathsf{A}(x)=\left(\frac{\Delta_1}{1-S_p}+i\right)\sqrt{\frac{2(S_p-S)}{C_3}}{{\rm sech}\left(\sqrt{\frac{S_p-S}{-C_1}}x\right)}.
\end{equation}

Figure~\ref{fig2} shows the real and imaginary parts of the weakly nonlinear solution (\ref{bump}) for $(\Delta_1,\eta_2)=(-4,0.01)$, $S_p-S=0.01$, and a large domain $l=200$ (see solid blue line). The orange dots show the exact numerical solution computed by means of a Newton-Raphson algorithm taking the profile defined by Eq.~(\ref{bump}) as initial solution guess. As shown in this plot, the agreement is excellent.

Equation~(\ref{normal_form}) describes locally the time-independent behavior of the system close to the pitchfork bifurcation, where the approximate asymptotic LS (\ref{bump}) fits well the exact solution. However, the good agreement shown in Fig.~\ref{fig2} worsens as $S$ moves away from $S_p$, and the systems enters the highly nonlinear regime.

\section{Localized states in the uniform-bistability scenario}\label{sec:4}
To unveil the features and bifurcation structure of LSs in the highly nonlinear regime 
(i.e., far apart from $S_p$) numerical approaches must be taken into consideration.
Here, we apply numerical path-continuation methods, based on a pseudo-archlength predictor and a Newton-Raphson corrector, to Eq.~(\ref{sta_normalized}). These methods allow us to track any time-independent state, either stable or unstable, as a function of the different parameters of the system. The computation of their stability is performed simultaneously by solving Eq.~(\ref{eigen}).

In this section, we focus on the bifurcation structure of LSs arising in the uniform-bistable configuration.
To begin the numerical continuation a suitable initial guess is necessary, which in our case, is provided by the asymptotic solution (\ref{bump}).
The outcome of this computation is shown in the bifurcation diagram of Fig.~\ref{fig3} for $\Delta_1=-4$, where solid (dashed) lines represent stable (unstable) states. Close to $S_p$, the LS is similar to the one depicted in Fig.~\ref{fig2}, and temporally unstable. Decreasing $S$ the LS increases its amplitude as it withdraws from $S_p$ and the system enters in a highly nonlinear regime. The LS stabilizes at the first left fold of the bifurcation curve which corresponds to a saddle-node bifurcation SN$_1^l$. The stable high amplitude LS here is similar to the one depicted in Fig.~\ref{fig3}(i), and their Fourier transform $\mathcal{F}$ [i.e. $FT(\cdot)=10{\rm log}_{10}(|\mathcal{F}[\cdot]|^2)$] is depicted in Fig.~\ref{fig3}(i'). Its region of existence extends all the way until SN$_1^r$ where it loses its stability once more. Soon after passing this fold, a dip start to nucleate at $x=0$ which deepens with decreasing $S$. At SN$_2^l$ such state becomes stable and looks like the one depicted in Fig.~\ref{fig3}(ii).

Proceeding up in the bifurcation curve (i.e., increasing $||\mathsf{A}||^2$), the dip nucleation process continue and the LS broadens as more and more spatial oscillations (i.e., dips) appear at the top part of the structure, which now can be seen as an almost flat plateau around $\mathsf{A}_h^+$ embedded in $\mathsf{A}_h^0$. Several examples of such states along the bifurcation diagram and their Fourier transforms are shown in Figs.~\ref{fig3}(iii)-(vi) and \ref{fig3}(iii')-(vi').

At this stage the formation of the LSs [see Fig.~\ref{fig3}(vi)] is mediated by the locking of plane fronts like those shown in Fig.~\ref{fig1}(e). The damped oscillatory shape of the bifurcation curve and the stability of the LSs is a direct consequence of the plane front interaction and their locking around the uniform Maxwell point of the system $S_M$ \cite{parra-rivas_localized_2019}. The exponential weakening of the interaction as the fronts separate from one another results in a exponential shrink of the extension of LS solution branches with increasing $||\mathsf{A}||^2$, and therefore, to the collapse of the bifurcation curve to $S_M$. Due to this characteristic, the bifurcation structure depicted in Fig.~\ref{fig3} is commonly known as {\it collapsed homoclinic snaking} \cite{knobloch_homoclinic_2005-1,yochelis_reciprocal_2006,parra-rivas_dark_2016,parra-rivas_localized_2019}.

\begin{figure}[t]
	\centering
	\includegraphics[scale=1]{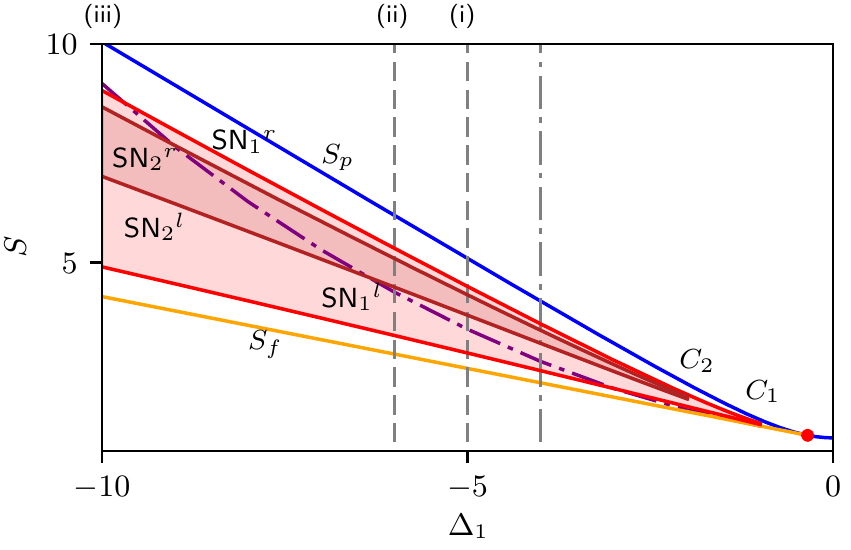}
	\includegraphics[scale=1]{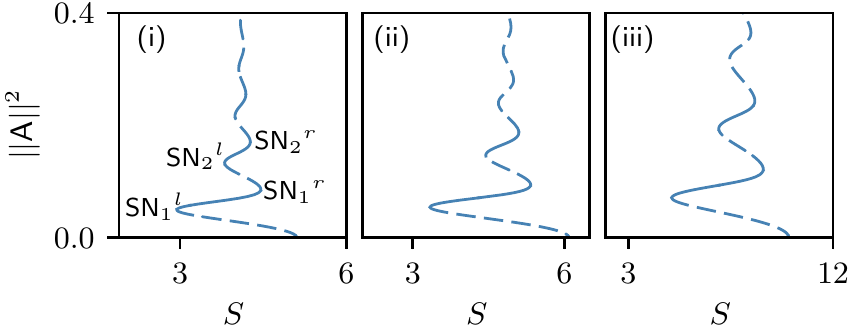}
	\caption{Phase diagram in the $(\Delta_1,S)$-parameter space showing the main attractor regions and bifurcations of the system in a uniform-bistable configuration for $\eta_2=0.01$. The vertical point-dashed line at $\Delta_1$ represents the bifurcation diagram shown in Fig.~\ref{fig3}. The different bifurcation lines correspond to the pitchfork bifurcation $S_p$, the saddle-node of $\mathsf{A}_h$ SN$_h$, the MI (see purple point-dashed line) and the saddle-node bifurcations SN$_1^{l,r}$ and SN$_2^{l,r}$ of the single- and two-bumps LSs. The bifurcation diagrams shown in sub-panels (i)-(iii) correspond to slices of constant $\Delta_1$ (namely $\Delta_1=-5,-6$ and $-10$) of the $(\Delta_1,S)$-phase shown on top. Solid (dashed) lines represent stable (unstable) states. The uniform nascent bistability point is marked with ${\color{red}\bullet}$.}
	\label{fig4}
\end{figure}

The collapsed snaking structure persists for different values of $\Delta_1$ as shown in the $(\Delta_1,S)$-phase diagram of Fig.~\ref{fig4}. This diagram shows the main bifurcation lines of the system for this regime of parameters: the stationary uniform bifurcation lines $S_p$ and $S_f$, the MI line $S_T^+$, and the first four folds SN$_{1,2}^{l,r}$ of the collapsed snaking diagram shown in Fig.~\ref{fig3}. This last diagram corresponds to a slice of constant $\Delta_1$ (see dashed vertical line) of Fig.~\ref{fig4}.

Increasing $|\Delta_1|$, the existence region of the single-peak LS (see shadowed region in-between SN$_1^{l,r}$) broadens as it does the region of existence of the multi-bumps LSs, here spanned by the lines SN$_2^{l,r}$. Decreasing $\Delta_1$, however, SN$_2^{l,r}$ approach one another and eventually collide in a cusp bifurcation $C_2$ where single-dip states disappear. Similarly, the region of existence of the single-peak LS shrinks as SN$_1^{l,r}$ gradually come closer, and eventually, it also fades away at another cusp $C_1$. 

The cusp bifurcations involving the collision of SN$_{i}^{l,r}$ (with $i>2$) occur successively in a similar fashion than in Ref.~\cite{parra-rivas_dark_2016}. The single-peak LSs persist below the MI curve $S_T^+$, and therefore outside the uniform bistability region, in contrast to the multi-bump states which need the presence of bistability to be formed.

For the range of parameters studied here, the single-bump and two-bumps LSs remain stable in-between SN$_1^{l,r}$ and SN$_2^{l,r}$ respectively. However, multi-bumps states with more than two-bumps lose their stability when increasing the value of $|\Delta_1|$. The loss of stability can be clearly observed in the bifurcation diagram shown in Figs.~\ref{fig4}(i)-(iii).

\section{Breathers in the uniform-bistability scenario}\label{sec:4b}
\begin{figure}[!t]
	\centering
	\includegraphics[scale=1]{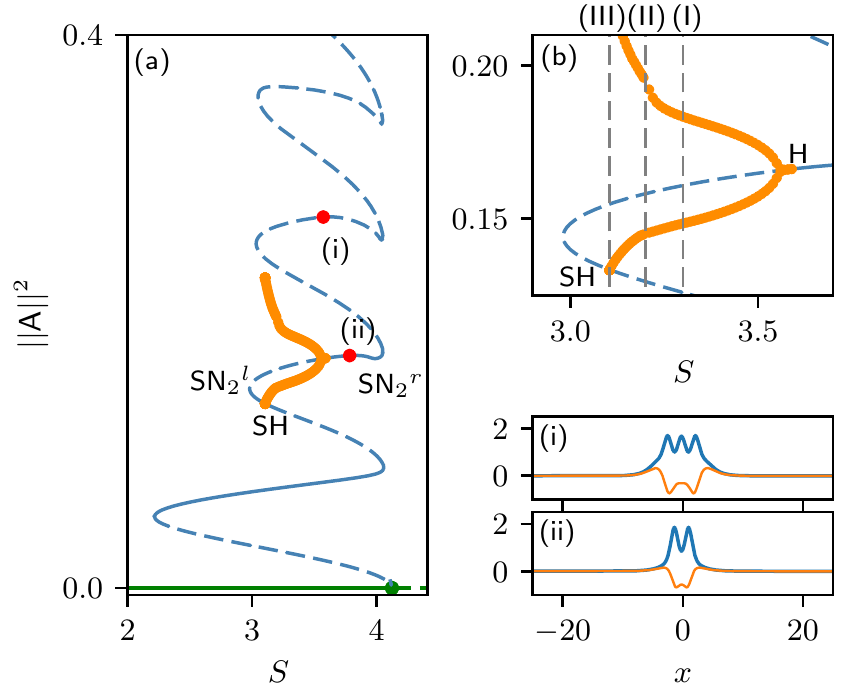}
	\includegraphics[scale=1]{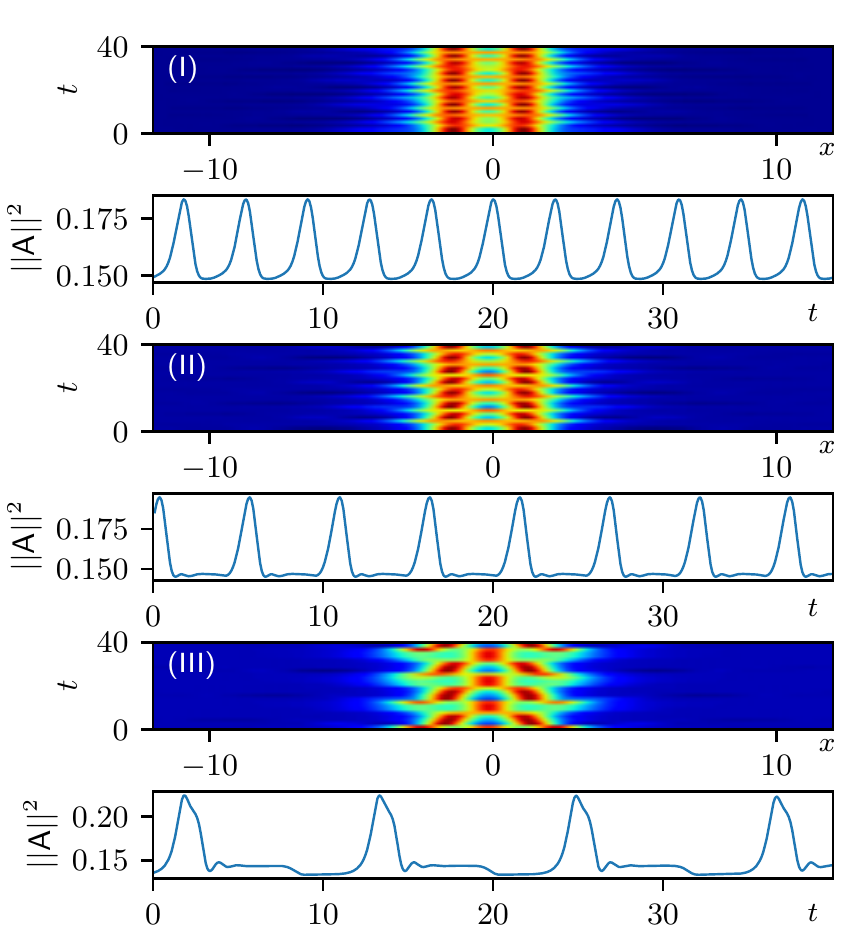}
	\caption{
		 Panel (a) shows the bifurcation diagram showing $||\mathsf{A}||^2$ as a function of $S$ for $\eta_2=0.25$ and $\Delta_1=-4$. Stable (unstable) solution branches are plotted using solid (dashed) lines. The orange dots show the variation of the maxima and minima of the breather LS from their birth at a supercritical Hopf bifurcation to their disappearance at the SH bifurcation. A close-up view of these branches are plotted in panel (b). Labels (i)-(ii) correspond to the static LSs shown on the right. The panels  (I)-(III) shown below correspond to the vertical dashed lines in (b), and represent the spatiotemporal evolution of the breathers [top] and the temporal evolution of its norm [bottom].}
	\label{fig7a}
\end{figure}

So far, the states reported here were static, i.e., they do not show any kind of permanent temporal dynamics.  However, a careful exploration of the parameter space, applying direct numerical simulations, shows that {\it spatially localized oscillations} can also arise when modifying the pump GVD parameter $\eta_2$. Figure~\ref{fig7a} shows the bifurcation diagram of the LSs in a uniform-bistable regime for $\eta_2=0.25$ and $\Delta_1=-4$, where the $L_2$-norm $||\mathsf{A}||^2$ is plotted as a function of $S$. Here, similarly to the bifurcation diagrams shown in Figs.~\ref{fig4}(i)-(iii), LSs with more than two bumps are temporally unstable [see for example Fig.~\ref{fig7a}(i) corresponding to the red dot shown in Fig.~\ref{fig7a}(a)].

\begin{figure}[!t]
	\centering
	\includegraphics[scale=1]{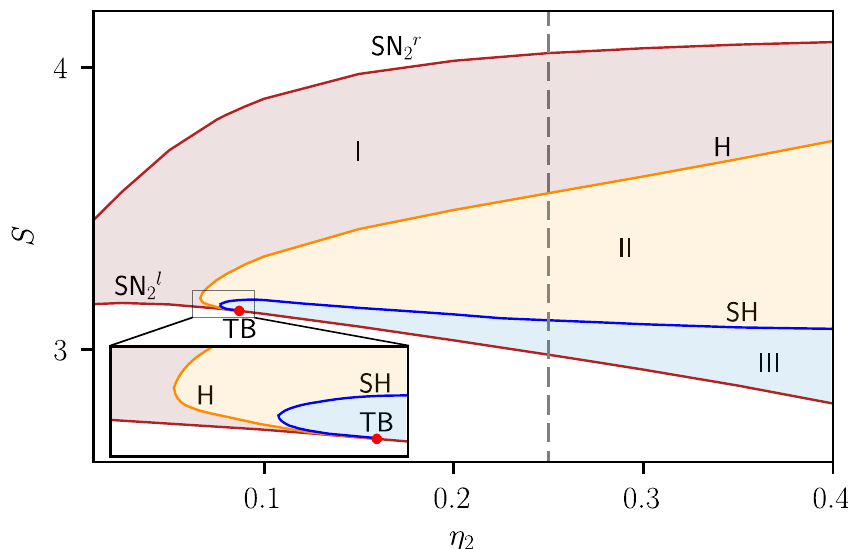}
	\caption{Phase diagram in the $(\eta_2,S)$-parameter space associated with the two-bump LSs showing the bifurcation lines SN$_2^{l,r}$, H and SH, and the different dynamical regions. The temporal dynamics arise from the TB point (see red dot) as depicted in the inset close-up view. In the shadowed red region I the LSs are static. Breathers exist in region II between the H and SH lines. In region III the two-bump LS is unstable and breathing behavior absent.		
 }
	\label{fig7b}
\end{figure}
Decreasing $||{\mathsf A}||^2$, the two-bump (i.e., single-dip) LS [see profile in Fig.~\ref{fig7a}(ii)] is stable when it arises from SN$_2^r$ and destabilizes at a supercritical Hopf (H) bifurcation. Below this point, oscillatory LSs, also called {\it oscillons} or {\it breathers}, emerge.
One example of the periodic temporal evolution of such state is shown for $S=3.3$ in Fig.~\ref{fig7a}(I)[top] together with the temporal variation of its norm [bottom]. The maxima and minima of the breather's norm are plotted in Fig.~\ref{fig7a}(a) and Fig.~\ref{fig7a}(b) [close-up view of (a)]
using orange dots.

Decreasing $S$, the amplitude and period of the breather increase. An example of such situation is shown in Fig.~\ref{fig7a}(II) for $S=3.2$ [see the dashed vertical lines shown in Fig.~\ref{fig7a}(b)]. Decreasing $S$ even further, the breather eventually collides with the two-bump unstable LS [see Fig.~\ref{fig7a}(b)] and is destroyed in a Shilnikov homoclinic (SH) bifurcation \cite{glendinning_stability_1994,glendinning_local_1984,gaspard_bifurcation_1984}. Approaching this global bifurcation, the breathers' oscillation period diverges with a  characteristic scaling law $T\propto{\rm ln}(S-S_{\rm SH})$ \cite{glendinning_stability_1994} [not shown here]. An example of a breather and its oscillation period very close to this bifurcation is shown in Fig.~\ref{fig7a}(III) for $S=3.10435$. Very close to this point, the system may exhibit excitability as has been reported in other optical systems \cite{gomila_excitability_2005}.



The $(\eta_2,S)$-phase diagram plotted in Fig.~\ref{fig7b} shows the bifurcation lines SN$_2^{l,r}$, H and SH for $\Delta_1=-4$. The vertical dashed line corresponds to the bifurcation diagram shown in Fig.~\ref{fig7a}.
To the left of the H line (see red shadowed region I) the two-bump LSs are stable. The breather LS, created at H, persist until the SH bifurcation line (see shadowed orange region II) where it is destroyed. Below this line (see region III), the breathing behavior is absent, and the two-bump LSs are unstable. Here the temporal evolution of the system leads to  the closest attractor of the system, which corresponds to the single-bump LS.


Decreasing $\eta_2$, the H bifurcation line eventually folds back and collides with SN$_2^l$ in a Takens-Bogdanov (TB) codimension-two bifurcation \cite{guckenheimer_nonlinear_1983,kuznetsov_elements_2004}, which is also responsible for the emergence of the SH line tangent to H. The birth of H and SH lines from the TB point is shown in the close-up view of Fig.~\ref{fig7b}.


\section{Localized states in the Turing-bistability scenario}\label{sec:5}
\begin{figure*}[t]
	\centering
	\includegraphics[scale=1]{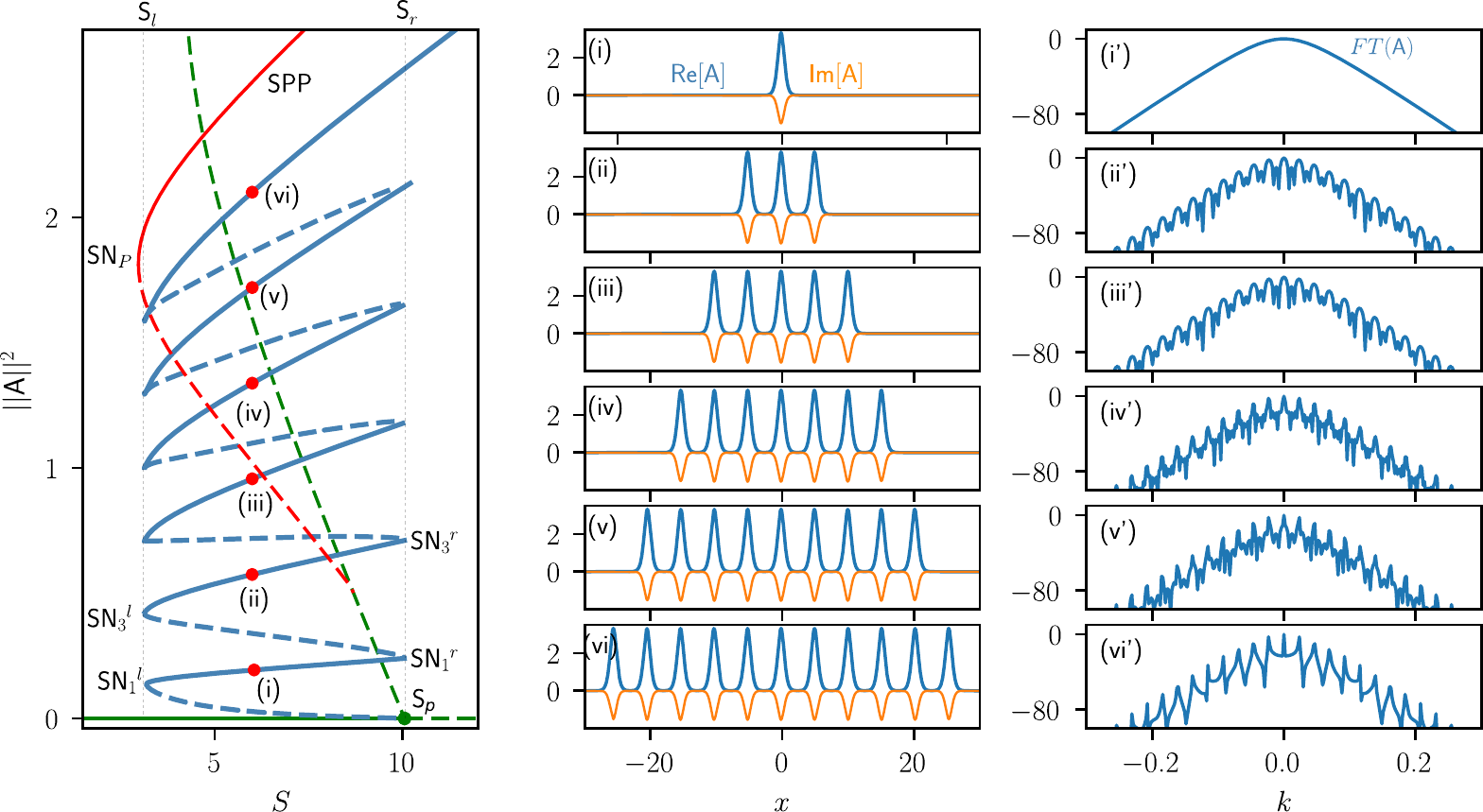}
	\caption{
	Standard homoclinic snaking in a Turing-bistable configuration. This bifurcation curve arises from $S_p$ at $\mathsf{A}_h^0$, and snakes in $S$ with a constant amplitude within the bistable region. Eventually, the diagram connects with a subcritical periodic Turing pattern emerging from $\mathsf{A}_h^-$. Panels (i)-(vi) show different LS solutions along this curve. Panels (i')-(vi') represent the frequency spectrum of the LSs. For the computations we have considered $l=60$, and the parameters are fixed to $(\Delta_1,\eta_2)=(-10,1)$.}
	\label{fig5}
\end{figure*}

In this section we follow a similar procedure to the one shown in Sec.~\ref{sec:4}, focusing this time on the Turing-bistable configuration shown in Fig.~\ref{fig1}(c). The outcome of the path-continuation procedure, starting from the initial analytical guess (\ref{bump}), leads to the bifurcation diagram shown in  Fig.~\ref{fig5}. As in the uniform-bistability scenario, the single-peak LS [see Fig.~\ref{fig5}(i)] increases its amplitude with decreasing $S$, stabilizes at SN$_1^l$ and persists stably until SN$_1^r$. Beyond this fold, however, the LSs undergo a bifurcation structure completely different to the collapsed homoclinic snaking depicted in Fig.~\ref{fig3}. Here, the LSs solution curve (see lines in blue) oscillates back-and-forth
within the locking, or snaking, region $S_l<S<S_r$ \cite{woods_heteroclinic_1999,burke_snakes_2007} which extends to the whole Turing-bistable region $S_P<S<S_p$.
This oscillation maintains a constant amplitude all along the diagram and reflects the successive addition of a pair of Turing pattern's peaks, one on each side of the state, as one follows the diagram upwards (i.e. increasing energy $||\mathsf{A}||^2$). The symmetric nucleation of a pair of peaks can be easily observed in the profiles shown in Fig.~\ref{fig5}(i)-(vi). The LS states belonging to this curve are composed by an {\it odd} number of peaks, and hereafter we refer to them as $\Gamma_o$. The different folds of the curve correspond to saddle-node bifurcations that we have labeled SN$_i^{l,r}$, with $i$ corresponding to the number of peaks involved in the structure. At each saddle-node, the LSs gain and lose stability in a similar manner than in the collapsed snaking scenario.

In a finite domain like ours, the nucleation process ends, once the domain is completely filled. The LSs solution curve then reconnects with one of the multiple subcritical SPPs within the bistability region as shown in Fig.~\ref{fig5}. 

This bifurcation scenario shares most features with the {\it standard homoclinic snaking} \cite{woods_heteroclinic_1999}, and in what follows we refer to it using the same term. In that context, the formation of LSs is related with a complex process known as {\it heteroclinic tangle} \cite{woods_heteroclinic_1999,gomila_bifurcation_2007}, which can be also understood through a locking mechanism involving patterned fronts. Thus, both LSs and their bifurcation structure can be predicted completely from the underlying bifurcation structure of the patterned fronts \cite{beck_snakes_2009,makrides_predicting_2014}. 

In the standard homoclinic snaking, $\Gamma_o$ emerge from the Turing instability together with the subcritical spatially periodic pattern, and another family of states characterized by LSs with an even number of pattern peaks (hereafter $\Gamma_e$). Here, in contrast, no Turing instability is present, and the snaking curve $\Gamma_o$ arise from the pitchfork bifurcation at $S_p$, while the periodic pattern emerges from one of the many branching bifurcation points occurring along $\mathsf{A}_h^-$ [see Fig.~\ref{fig5}]. Furthermore, the normal form (\ref{normal_form}) around the $S=S_p$ only supports single pulse solutions [see Eq.~(\ref{bump})], and thus $\Gamma_e$, if it exists, can not emerge from that point. 
\begin{figure}[!t]
	\centering
	\includegraphics[scale=1]{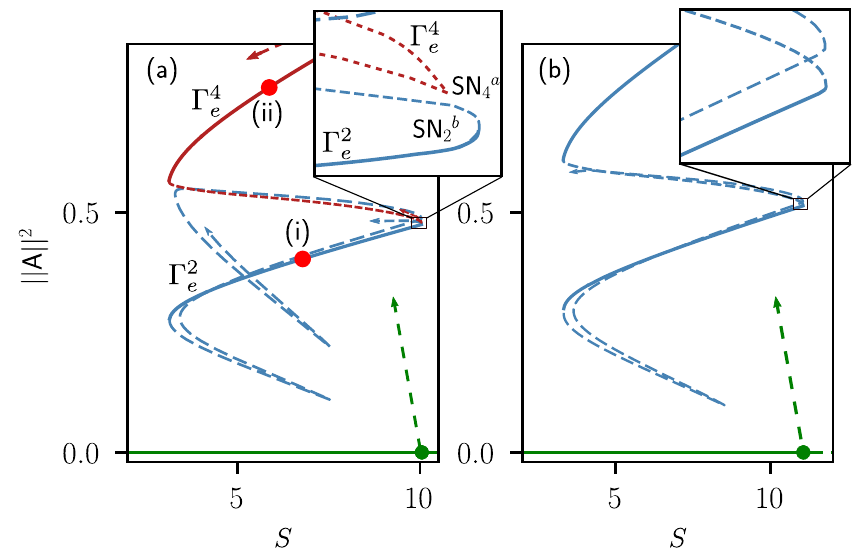}
	\includegraphics[scale=1]{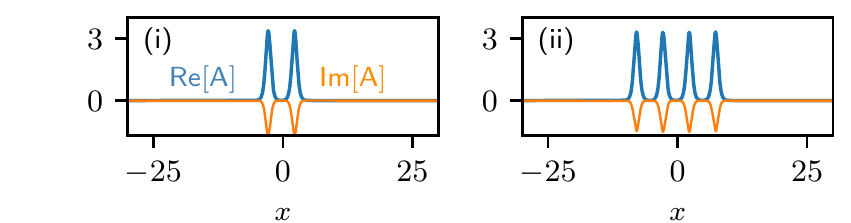}
	\caption{(a) Solution state curves $\Gamma_e^2$ and $\Gamma_e^4$ for $\Delta_1=-10$. The close-up view shows the saddle-node bifurcations SN$_2^b$ and SN$_4^a$. The labels (i)-(ii) correspond to the LSs profiles shown below. (b) Solution state curve $\Gamma_e$ for $\Delta_1=-11$, after the reconnection of SN$_2^b$ and SN$_4^a$ through a necking bifurcation. }
	\label{fig6b}
\end{figure}

At this stage, one question to answer is whether $\Gamma_e$ states exit, and if so, how do they organize. To answer this question we first integrate numerically Eq.~(\ref{normalized}) starting from a suitable initial condition, which allows us to compute $\Gamma_e$ LSs of different extensions. Afterwards, we can numerically continue such states in $S$, computing their bifurcation curves. The results of these computations are shown in Fig.~\ref{fig6b}(a) for the same parameters than the diagram shown in Fig.~\ref{fig5} (i.e., $\Delta_1=-10$). For this set of parameters, states with different (even) peak numbers are disconnected from each other, and organized in 
several bifurcation curves that we label $\Gamma_e^i$, with $i$ denoting the number of peaks of each state. Figure~\ref{fig6b}(a) shows $\Gamma_e^2$ and $\Gamma_e^4$, and two examples of stable LSs are plotted in Fig.~\ref{fig6b}(i)-(ii). $\Gamma_e^2$ is detached from ${\mathsf A}_h^0$, confirming our previous hypothesis. A detailed examination of these curves [see close-up view in Fig.~\ref{fig6b}(a)] shows that $\Gamma_e^2$ and $\Gamma_e^4$ are disconnected and undergo two saddle-nodes SN$_2^b$ and SN$_4^a$, which are very close to one another.
Increasing the energy, similar disjoint curves are found involving LSs with $6,8,10,\dots$ peaks (not shown here). 

For larger values of $|\Delta_1|$, SN$_2^b$ and SN$_4^a$ eventually collide in a necking bifurcation \cite{prat_12_2002}, and $\Gamma_e^{2,4}$ merge in a single one, as depicted in Fig.~\ref{fig6b}(b) for $\Delta_1=-11$. Increasing $|\Delta_1|$ even further, similar reconnections occur between wider states, and eventually, a fully connected bifurcation curve $\Gamma_e$ appears.

\begin{figure}[!t]
	\centering
	\includegraphics[scale=1]{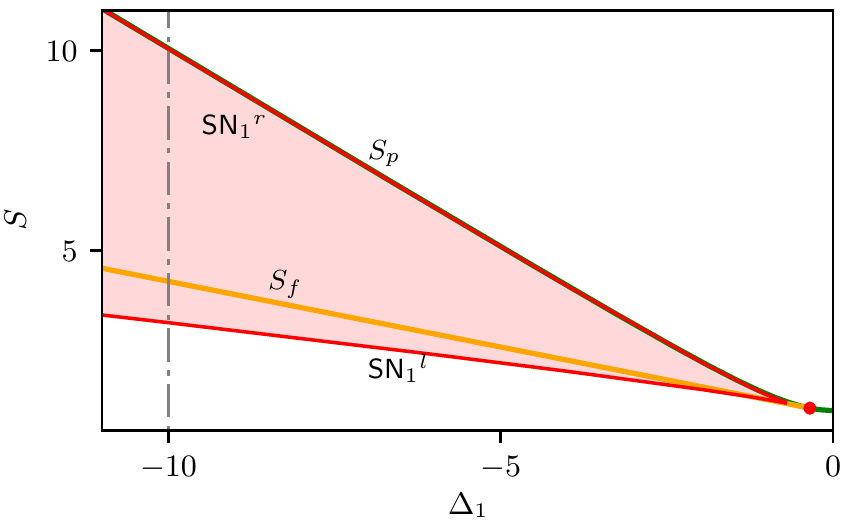}
	\caption{Phase diagram in the $(\Delta_1,S)$-parameter space showing the main attractor regions and bifurcations of the system in a Turing-bistable configuration for $\eta_2=1$. The vertical point-dashed line at $\Delta_1=-10$ corresponds the bifurcation diagram shown in Fig.~\ref{fig5}. The different bifurcation branches are the pitchfork bifurcation $S_p$, the saddle-node of $\mathsf{A}_h$ (SN$_h$), and the saddle-node bifurcations SN$_1^{l,r}$ in red. Here, SN$_3^{l,r}$ overlap with SN$_1^{l,r}$. The red shadowed region corresponds to the snaking region. The uniform nascent bistability point is marked with ${\color{red}\bullet}$.}
	\label{fig6}
\end{figure}
These localized patterns persist while changing the value of $\Delta_1$ as shown in the $(\Delta_1,S)$-phase-diagram of Fig.~\ref{fig6}. This diagram shows the main bifurcation of the system, namely $S_p$, $S_f$, SN$_{1}^{l,r}$ and SN$_{3}^{l,r}$, together with the pinning region (see shadowed area).
Similarly to the collapsed snaking case (see Fig.~\ref{fig4}), the localization region enlarges with increasing $|\Delta_1|$ and shrinks with decreasing it, until eventually the pairs of saddle-node bifurcations SN$_i^{l,r}$ collide in sequence of cusp bifurcations, and the LSs disappear sequentially as one approaches the uniform nascent bistability point at $\Delta_1=1/\beta$.  

Despite the persistence of these states, the bifurcation structure shown in Fig.~\ref{fig5} is not preserved and suffers a disconnection process, mediated by necking bifurcations while decreasing $\Delta_1$, similar to the one shown in Fig.~\ref{fig6b}.

As far as we know, this type of homoclinic snaking is rare, and so far has only been previously reported in the context of the Legendre-Lefever model describing  vegetation patterns in dryland ecosystems \cite{zelnik_desertification_2017}. In that case, however, $\Gamma_o$ emerge from a transcritical bifurcation at the trivial state.
Moreover, the reconnection of bifurcation curves through necking bifurcation is an infrequent phenomenon in the context of homoclinic snaking. However, we believe that it can be found in other systems sharing the same type of features \cite{zelnik_desertification_2017}.

\section{Implications of the walk-off: propagating localized states}\label{sec:6}

\begin{figure}[!t]
	\centering
	\includegraphics[scale=1]{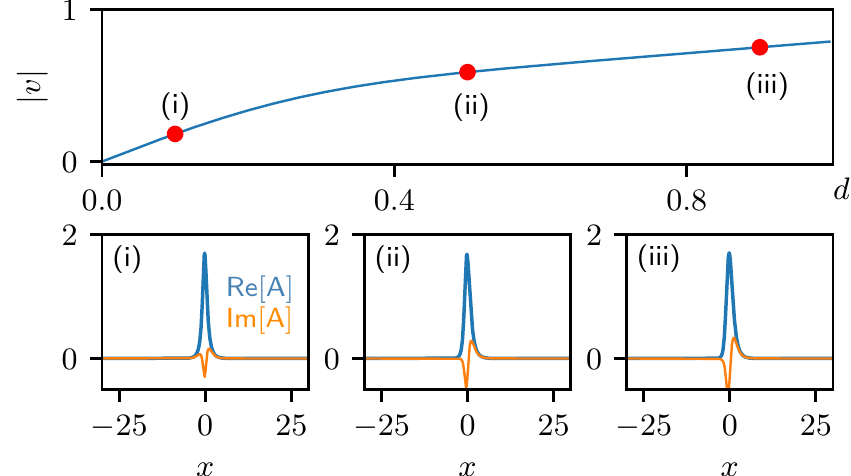}
	\caption{Dependence of $|v|$ with $d$ for a single-peak LS. The asymmetry of these states increase with $d$ as shown in panels (i)-(iii). Here $(\eta_2,\Delta_1,S)=(0.01,-4,3)$.}
	\label{fig70}
\end{figure}

In the previous sections we have neglected the effect of group velocity mismatch or temporal walk-off by setting $d=0$. In the presence of walk-off ($d\neq0$), the reflection symmetry $x\rightarrow-x$ is broken, and the LSs studied previously start to propagate at a constant speed which depends on $d$. An example of this dependence, computed through path-numerical continuation in $d$, is shown in Fig.~\ref{fig70} for the single-peak LS with $(\eta_2,\Delta_1,S)=(0.01,-4,3)$. Increasing $d$, the $x\rightarrow -x$ asymmetry increases and so does the speed $|v|$ of the LS [see profiles (i)-(iii)].
In what follows we focus on a weak reflection symmetry breaking $(d=0.1)$, and present the implication that such asymmetry may have on the bifurcation structures of the LSs studied previously.

\begin{figure*}[t]
	\centering
	\includegraphics[scale=1]{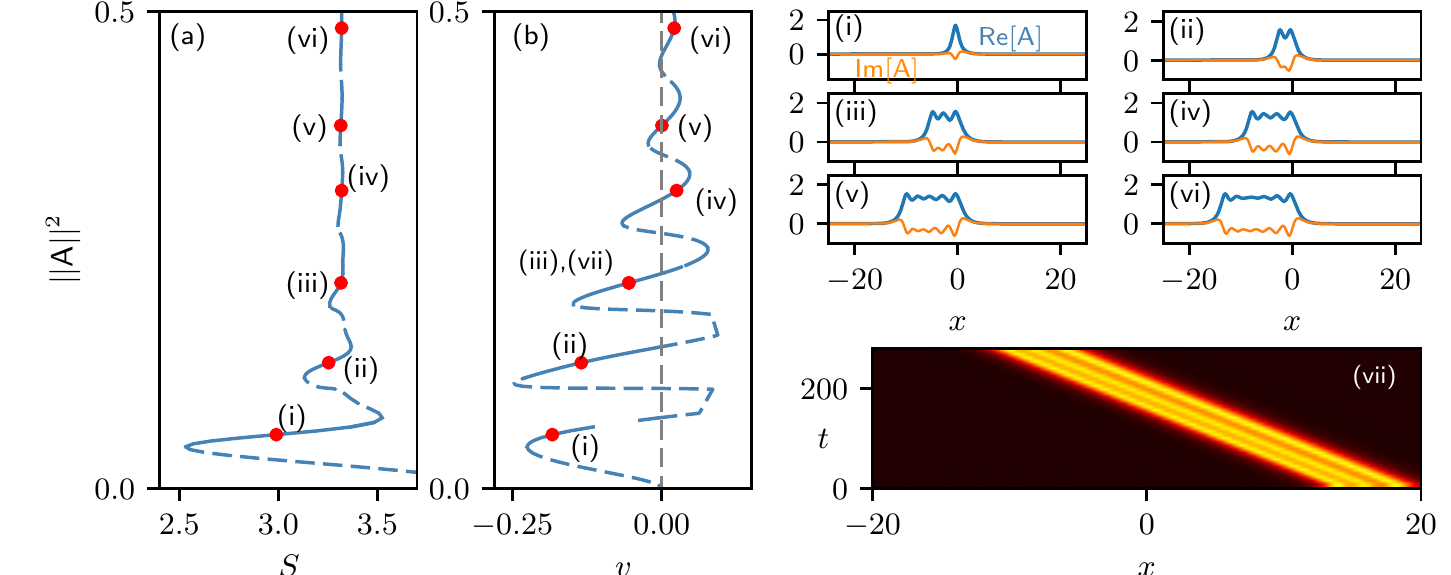}
	\caption{Implication of walk-off on LSs in the uniform-bistable scenario for $d=0.1$. Panel (a) shows the modification of the collapsed snaking shown in Fig.~\ref{fig3}, whereas panel (b) shows the modification of the speed of the LSs with the width measured through the norm $||\mathsf{A}||^2$. Panels (i)-(vii) show several examples of asymmetric and propagating LSs along the diagrams shown in (a) and (b). Panels (vii)-(viii) show the propagation with different speeds of two 3-bumps state. Here, 
		$(\eta_2,\Delta_1,\varrho)=(0.01,-4,-4)$.}
	\label{fig7}
\end{figure*}

\begin{figure*}[t]
	\centering
	\includegraphics[scale=1]{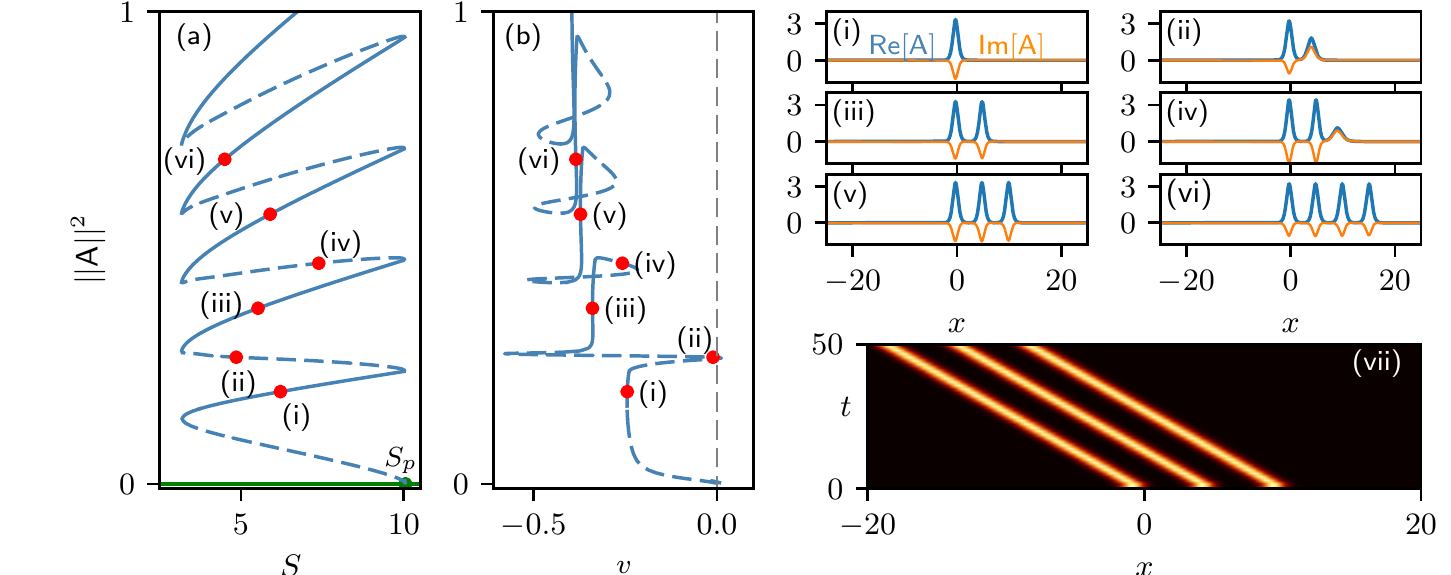}
	\caption{Implication of walk-off on LSs in the Turing-bistable scenario for $d=0.1$. Panel (a) shows the modification of the standard homoclinic snaking shown in Fig.~\ref{fig5}, whereas panel (b) shows the modification of the speed of the LSs with their width, measured through the norm $||\mathsf{A}||^2$. Panels (i)-(vii) show several examples of asymmetric and propagating LSs along the diagrams shown in (a) and (b). Panels (vii)-(viii) show the propagation of two different states. Here,$(\eta_2,\Delta_1,\varrho)=(1,-10,-4)$.}
	\label{fig8}
\end{figure*}
Figure~\ref{fig7} shows the effect of a weak walk-off on the collapsed snaking presented in Fig.~\ref{fig3}. For this value of $d$, the collapsed snaking structure is preserved [see Fig.~\ref{fig7}(a)], up to small shape modifications due to the effect of the asymmetry on the front's interaction. Panels~\ref{fig7}(i)-(vi) show some examples of propagating LSs along the bifurcation diagram shown in Fig.~\ref{fig7}(a), where we can easily appreciate their asymmetry in $x$. 

The path-continuation algorithm used here allows us to also compute the speed $v$ of these states as a function of $S$ or any other tracking parameter. The diagram plotted in Fig.~\ref{fig7}(b) shows $v$ as a function of the width of the LS, here measured through its $L_2$-norm $||\mathsf{A}||^2$. Proceeding up in the diagram, $v$ oscillates in an irregular manner around $v=0$, and broad states propagate with positive speed. This feature agrees with the case previously reported regarding the influence of third-order dispersion on dark LSs in Kerr cavities \cite{parra-rivas_coexistence_2017}. The propagation of a 3-bumps LSs with negative velocity is shown in Fig.~\ref{fig7}(vii) for $S\approx3.317$.   


We have also studied the effect of the walk-off on LSs in a standard homoclinic snaking context. The results are presented in Fig.~\ref{fig8}. In this case, the $x-$reflection symmetry breaking yields the merging of LSs with odd and even number of peaks, and thus to the {\it mixed snaking} shown in Fig.~\ref{fig8}(a) which alternates odd and even LSs. Examples are depicted in Figs.~\ref{fig8}(i)-(vi). For this value of $d$, the LSs profiles look quite symmetric. However, all these states drift with a negative speed whose dependence with $||\mathsf{A}||^2$ is shown in Fig.~\ref{fig8}(b). Here, the speed along a given stable segment of the diagram seems almost constant. Furthermore, $|v|$ increases with the width (i.e., with $||\mathsf{A}||^2$) of each LS and collapses to a vertical asymptote, such that the LSs shown in Fig.~\ref{fig8}(v)-(vi) have almost the same speed.

In each right fold of these diagrams a single new peak is nucleated at the right of the state as depicted in the LSs profiles (ii) and (iv). The emergence and growing of this new peak yields a large variation of the speed [see Fig.~\ref{fig8}(b)].
Similar scenarios have also been studied in other conservative and dissipative non-spatial reversible systems \cite{sandstede_snakes_2012,makrides_predicting_2014,parra-rivas_third-order_2014}. Other plausible scenario that could appear in this context relates with the breaking of the homoclinic snaking in a stuck of isolas \cite{burke_swift-hohenberg_2009,sandstede_snakes_2012,parra-rivas_third-order_2014}. However, for the range of parameters studied here, such scenario is absent.

\section{Discussion and conclusions}\label{sec:8}
The existence of single-peak LSs in phase mismatched singly resonant dispersive DOPOs was first reported in Ref.~\cite{nie_quadratic_2020}, although a detailed taxonomic classification of the different types of LSs was so far lacking.

In this work we have presented a complete and detailed description of the bifurcation structure, stability and dynamics of the variety of LSs, either static or dynamic, arising in the context of singly resonant DOPOs.
In the mean-field approximation, these cavities can be described by parametrically forced Ginzburg-Landau equation (see Sec.~\ref{sec:1}) with nonlocal nonlinearity $A^2\otimes\mathsf{J}$ describing a time delay induced by the pump field parameters through the kernel or response function $\mathsf{J}$ \cite{mosca_modulation_2018}. The formation and bifurcation structure of LSs in systems with nonlocal long-range coupling terms appear naturally in a variety of scientific domains, and has been particularly tackled in the context of neural field models \cite{rankin_continuation_2014,schmidt_bumps_2020}, nonlinear optics \cite{parra-rivas_influence_2021}, and in the prototypical Swift-Hohenberg equation \cite{morgan_swifthohenberg_2014}.

We have applied a pattern forming and dynamical systems approach \cite{cross_pattern_1993}, which has allowed us to unveil the general mechanism behind the formation of LSs and their bifurcation origin. 
As usual when considering this procedure, the starting point is the analysis of the uniform or homogeneous steady state solution and its linear stability (Sec.~\ref{sec:2}). This analysis has allowed as to identify two regimes leading to the emergence of LSs: the {\it uniform-bistable} and the {\it Turing-bistable} regimes. In the first one, two uniform steady states coexist in a stable manner within the same interval of parameters. This situation allows the formation of plane fronts connecting such states, which eventually can lock, leading to LSs of different extensions. In the second case, in contrast, the coexistence appears between an uniform state and a spatially extended pattern. Similarly, patterned fronts may form and lock, leading to different kinds of localized patterns. 

We have shown, applying weakly nonlinear analysis (Sec.~\ref{sec:3}), that in both scenarios small amplitude LSs with a single bump emerge from the pitchfork bifurcation undergone by the trivial uniform state $\mathsf{A}_h^0$. These weakly nonlinear LSs undergo a very different bifurcation structure which depends on the type of bistability.

In the uniform bistability regime, these small amplitude states undergo collapsed homoclinic snaking around the uniform Maxwell point of the system, as they enter the highly nonlinear regime (see Sec.~\ref{sec:4}). All along this diagram the region of existence of the different states shrinks as a function of their width (i.e., measured by $||\mathsf{A}||^2$). We have shown that LSs of this kind persist in the $(\Delta_1,S)$-parameter space and that they disappear when approaching the nascent uniform bistability point at $\Delta_1=1/\beta$. This scenario is very similar to the one reported in the context of doubly resonant DOPOs where the uniform-bistable configuration was explored \cite{parra-rivas_localized_2019}. 

Modifying the GVD parameter $\eta_2$, the system undergoes oscillatory instabilities (see Sec~\ref{sec:4b}) and breathers emerge. These dynamical states arise supercritically from the stationary two-bump states and are destroyed in a Shilnikov homoclinic bifurcation. We have analyzed the modification of the different dynamical regions as a function of $\eta_2$. Furthermore, we have found that the birth and death of the breathing behavior is related with the presence of a TB codimension-two point, from where the H and SH bifurcations arise.
As far as we know, the single-bump stable LS does not suffer this kind of instability.

In the Turing-bistable scenario LSs undergo a particular bifurcation structure that shares similarities with the standard homoclinic snaking (see Sec.~\ref{sec:5}). Differently to the collapsed snaking case, here LSs of different width exist within the same interval of parameters, in the so called snaking region. The LSs and spatial pattern emerge from different points: while LSs with an odd number of peaks emerge from the pitchfork bifurcation at $S_p$, the pattern related with these states bifurcates from one of the many branching bifurcation points along $\mathsf{A}_h^-$. Localized patterns with an even number of peaks are also present, although their bifurcation curves are detached from $S_p$. This particular type of bifurcation structure is not that common, and has been also reported in the context of mathematical models describing vegetation patterns in dryland ecosystems \cite{zelnik_desertification_2017}. Localized patterns persist in the $(\Delta_1,S)$-parameter space, although their homoclinic snaking does not as it breaks up in different bifurcation curves through a sequence of necking bifurcations.  

Finally we have also elucidated how the bifurcation structure of LSs is modified when considering a weak walk-off ($d\ll1$) (see Sec.~\ref{sec:6}). In the uniform-bistable configuration the collapsed snaking is conserved, while in the Turing-bistable scenario branches of LSs with even and odd number of peak interconnect, forming a mixed snaking curve similar to the one reported in the context of Kerr cavities with third-order dispersion \cite{parra-rivas_third-order_2014}. In both cases, the dependence of the LSs speed with the energy $||\mathsf{A}||^2$ has been computed.
  
In this study we have considered a fixed phase mismatch $\varrho\approx -1.27\pi$ and positive GVD $\eta_{1,2}$. This election corresponds to the $\varrho\eta_2<0$ regime reported in \cite{nie_quadratic_2020}. Other regimes of parameters are so far unexplored and could lead to other interesting localization phenomena, as well as patterned dynamics that we expect to explore in the future. 

The bifurcation structure underlying the localization phenomena presented here establishes a useful map linking different kinds of LSs with the control parameters of singly resonant DOPOs which will prove hopefully useful for their experimental exploration.

\appendix

\section{Mean-fild normalization}\label{sec:A}
Under the mean-field approximation the dynamics of intracavity signal field in singly resonant dispersive DOPO is described by the following partial differential equation with nonlocal nonlinearity \cite{mosca_modulation_2018}
\begin{multline}\label{dimless}
t_R\partial_t A=-(\alpha_1+i\delta_1)A-i\frac{k_1''L}{2}\partial_\tau^2A\\-\mu^2\bar{A}(A^2\otimes J)+\mu{\rm sinc}(\varrho/2)B_{in}\bar{A}e^{i(\pi/2-\varrho/2)},
\end{multline}
 where $A$ is the envelope of the signal field, $t$ is the slow time describing time evolution of $A$ over successive round-trips, $t_R$ is the round-trip time, $\tau$ is the fast time describing the temporal profiles in the retarded time frame, $\alpha_1$ are the total linear cavity losses associated with the signal, and $\delta_1$ is the phase detuning to the closest frequency resonance, $k_1''$ is the group velocity dispersion (GVD) associated with $A$, and $L$ is the length of the cavity. $\mu=\kappa L$ with $\kappa$ being the second order nonlinear coefficient, $\varrho=\Delta k L$, where $\Delta k$ is the phase mismatch, and $B_{in}$ is the continuous wave driving field or pump. $A^2\otimes J$ represents the nonlocal nonlinearity, with $\otimes$ representing the convolution between $A^2$ and the nonlocal response function or kernel
\begin{equation}
J(\omega)=\frac{1}{2\pi}\int_{-\infty}^{\infty}j(\omega)e^{-i\omega \tau}d\tau,
\end{equation}
with 
\begin{align}
j(\omega)=\frac{1-e^{-iZ(\omega)}-iZ(\omega)}{Z(\omega)},	
\end{align}
\begin{align}
Z(\omega)=\varrho+ig(\omega)L,	
\end{align}
and
\begin{align}
g(\omega)=-\alpha_{c,2}/2+i\left(\Delta k'\omega+\frac{k_2''}{2}\omega^2\right).
\end{align}
The last expression contains the group velocity mismatch $\Delta k'$, the propagation losses $\alpha_{c,2}$, and the group velocity dispersion $k_2''$ associated with the pump field $B$, which is extracted from the cavity at each round-trip.
  
As done in similar studies \cite{leo_frequency-comb_2016,mosca_modulation_2018,nie_quadratic_2020}, we neglect the losses associated with the pump field, that is, we take $\alpha_{c,2}=0$. Thus, the nonlocal kernel splits as 
\begin{equation}
j(\omega)=j_R(\omega)+ij_I(\omega),
\end{equation}
with 
\begin{align*}
j_R(\omega)=\frac{1}{2}{\rm sinc}^2(Z(\omega)/2), && j_I(\omega)=\frac{{\rm sinc}(Z(\omega))-1}{Z(\omega)},
\end{align*}
and 
$Z(\omega)=\varrho-L\Delta k'\omega-\frac{1}{2}k_2''L\omega^2.$
\\

Considering the transformations
\begin{align*}
A=&A_ce^{i\psi}{\mathsf A}/\sqrt{a(\varrho)},&\tau=&\tau_c x,&t=&t_cT,
\end{align*}
with 
$a(\varrho)\equiv j_R(0)=\frac{1}{2}{\rm sinc}^2(\varrho/2)$, and the characteristic coefficients and phase
\begin{align*}
A_c=&\frac{\sqrt{\alpha_1}}{\mu},&t_c=&t_R/\alpha_1,&\tau_c=\sqrt{\frac{L|k_1''|}{2\alpha_1}},&&\psi=(\pi-\varrho)/4
\end{align*}
we can transform the dimensional Eq.~(\ref{dimless}) into the dimensionless equation
\begin{equation}\label{NLGL_normalized}
\partial_t {\mathsf A}=-(1+i\Delta_1)\mathsf{A}-i\eta_1\partial_{x}^2\mathsf{A}-\bar{\mathsf{A}}(\mathsf{A}^2\otimes\mathsf{J}) +S\bar{\mathsf{A}},
\end{equation}
where the normalized control parameters read
\begin{subequations}
	\begin{align}
	\Delta_1=\frac{\delta_1}{\alpha_1},&& S=\frac{\mu{\rm sinc}(\varrho/2)}{\alpha_1}B_{in},
	\end{align}

	\begin{align}
	\eta_1={\rm sign}(k_1''), &&\eta_2=\frac{\alpha_1k_2''}{|k_1''|},&& d=\sqrt{\frac{2\alpha_1L}{|k_1''|}}\Delta k'.
	\end{align}

\end{subequations}

\section{Weakly nonlinear analysis around the Pitchfork bifurcation}\label{sec:B}

Following~\cite{burke_classification_2008,parra-rivas_localized_2019} we fix $\Delta_1$ and consider the asymptotic expansion of the fields $U\equiv {\rm Re}[\mathsf{A}]$, and $V\equiv {\rm Im}[\mathsf{A}]$ as a function of the expansion parameter $\epsilon$ defined by $S=S_p+\delta\epsilon^2$, where $\delta$ is the bifurcation parameter. Then the expansion reads
\begin{equation}\label{expand}
\left[\begin{array}{c}
U \\ V
\end{array}
\right]=\epsilon\left[\begin{array}{c}
u_1 \\ v_1
\end{array}
\right]+\epsilon^3\left[\begin{array}{c}
u_3 \\ v_3
\end{array}
\right]+\cdots.
\end{equation}
where we allow each of the terms in the previous expansion to depend just on the long scale $X\equiv\epsilon x$. Considering Eq.~(\ref{expand}) the linear operator expands as 
\begin{equation}
\mathcal{L}=\mathcal{L}_0+\epsilon^2\mathcal{L}_2,
\end{equation}
with
\begin{subequations}
	\begin{equation}
	\mathcal{L}_0=\left[\begin{array}{cc}
	S_p-1 & \Delta_1 \\ -\Delta_1 &-(S_p+1)
	\end{array}
	\right],
	\end{equation}
	and
	\begin{equation}
	\mathcal{L}_2=\left[\begin{array}{cc}
	\delta & \eta_1\partial^2_{X} \\ -\eta_1\partial^2_{X} & -\delta
	\end{array}
	\right].
	\end{equation}
\end{subequations}
Similarly, the nonlinear operator becomes 
\begin{equation}
\mathcal{N}=\epsilon^2\mathcal{N}_2=-\left[\begin{array}{cc}
\mathcal{N}_2^a & \mathcal{N}_2^b\\
\mathcal{N}_2^b & -\mathcal{N}_2^a
\end{array}\right],
\end{equation}
with 
\begin{subequations}
	\begin{equation}
	\mathcal{N}_2^a=u_1^2\otimes \mathsf{J}_R-v_1^2\otimes \mathsf{J}_R-2u_1v_1\otimes \mathsf{J}_I  
	\end{equation}
	\begin{equation}
	\mathcal{N}_2^b= u_1^2\otimes \mathsf{J}_I-v_1^2\otimes \mathsf{J}_I+2u_1v_1\otimes \mathsf{J}_R.
	\end{equation}
\end{subequations}
The insertion of the previous expansions in the stationary equation (\ref{sta_normalized}) yields a hierarchy of equations for successive orders in $\epsilon$, which up to third order read: 
\begin{subequations}
	\begin{equation}\mathcal{O}(\epsilon): 
	\mathcal{L}_0\left[\begin{array}{c}
	u_1 \\ v_1
	\end{array}
	\right]=\left[\begin{array}{c}
	0 \\ 0
	\end{array}
	\right],		
	\end{equation}
	and 
	\begin{equation}
	\mathcal{O}(\epsilon^3):\mathcal{L}_0\left[\begin{array}{c}
	u_3 \\ v_3
	\end{array}
	\right]+(\mathcal{L}_2+\mathcal{N}_2)\left[\begin{array}{c}
	u_1\\v_1
	\end{array}
	\right]=\left[\begin{array}{c}
	0 \\ 0
	\end{array}
	\right]
	\end{equation}
\end{subequations}

At first order in $\epsilon$ the solvability condition provides
\begin{equation}
S_p=\sqrt{\Delta_1^2+1},
\end{equation}
which confirms the position of the pitchfork bifurcation already calculated in Sec.~\ref{sec:3}.
The solutions at this order are of the form 
\begin{equation}
\left[\begin{array}{c}
u_1 \\ v_1
\end{array}\right]=\left[\begin{array}{c}
\xi \\ 1
\end{array}\right]B(X),
\end{equation}
where $\xi=\Delta_1/(1-S_p)$ and $B(X)$ is the real envelope amplitude to be determined at next order in the expansion.

At $\mathcal{O}(\epsilon)$
\begin{equation}
\mathcal{L}_0\left[\begin{array}{c}
u_3 \\ v_3
\end{array}
\right]=-(\mathcal{L}_2+\mathcal{N}_2)\left[\begin{array}{c}
u_1 \\ v_1
\end{array}
\right].
\end{equation}

The amplitude equation about $S_p$ is then obtained from the solvability condition 
\begin{equation}
w^T\cdot\mathcal{L}_2\left[\begin{array}{c}
u_1 \\ v_1
\end{array}
\right]+w^T\cdot\mathcal{N}_2\left[\begin{array}{c}
u_1\\v_1
\end{array}\right]=\left[\begin{array}{c}
0\\0\end{array}\right],\end{equation}
where $w^T=[-\xi, 1]$, such that $\mathcal{L}^{\dagger}_0w=0$.	

The evaluation of the first term yields   
\begin{equation}
w^T\cdot\mathcal{L}_2\left[\begin{array}{c} u_1\\v_1
\end{array}\right]=-\delta(\xi^2+1) B-2\xi\eta_1\partial^2_{X} B,
\end{equation}
while the second one gives	
\begin{equation}
w^T\cdot\mathcal{N}_2\left[\begin{array}{c} u_1\\v_1
\end{array}\right]=(\xi^2+1)\mathcal{N}_2^aB(X).
\end{equation}
Thus, the only term of $\mathcal{N}_2$ contributing is $\mathcal{N}_2^a$. To evaluate this term we follow the same approach than in Refs.~\cite{nikolov_quadratic_2003,morgan_swifthohenberg_2014, kuehn_validity_2018}: 
we consider $B(X)$ to be constant within
its convolution with $\mathsf{J}$, using the implicit decoupling of the short and long length scales $x$ and $X=\epsilon x$ that
arises as $\epsilon\rightarrow0$.
With these considerations we obtain:
\begin{subequations}
	\begin{multline*}
	u_1^2\otimes J_R=\int_{-\infty}^{\infty} u_1^2(y)J_R(x-y)dy=\\ \xi^2\int_{-\infty}^{\infty} B(Y)^2J_R(x-y)dy\approx\xi^2 B(X)^2\int_{-\infty}^{\infty}J_R(x-y)dy
=\\\xi^2 B(X)^2j_R(0)=\xi^2 B(X)^2a(\varrho),
	\end{multline*}
	\begin{multline*}
	u_1v_1\otimes J_I=\int_{-\infty}^{\infty} u_1(y)v_1(y)J_I(x-y)dy\\\approx\xi B(X)^2\int_{\infty}^{-\infty}J_I(x-y)dy=\\
	\xi B(X)^2j_I(0)=\xi B(X)^2\beta(\varrho),
	\end{multline*}
\end{subequations}
\begin{subequations}
	and with the same approach
	\begin{equation*}
	v_1^2\otimes J_R\approx B(X)^2J_R(0)=B(X)^2a(\varrho),
	\end{equation*}		
\end{subequations}
Putting all together we have  
\begin{equation}
\mathcal{N}_2^a=(\xi^2-1)a(\varrho)-2\xi b(\varrho),
\end{equation}
which finally leads to the stationary amplitude equation 
\begin{equation}\label{normal_form_app}
C_1\partial^2_{X}B=\delta B+C_3B^3,
\end{equation}
with coefficients
\begin{align}
C_1\equiv\frac{-2\eta_1\xi}{1+\xi^2}, && C_3\equiv 2\xi \beta(\varrho)-(\xi^2-1).
\end{align}

\bibliographystyle{ieeetr}
\bibliography{LSs_general_OPO}

\end{document}